\documentclass[useAMS,usenatbib,onecolumn]{mn2e}
\usepackage{amssymb}
\usepackage{graphicx}

\title[Kelvin-Helmholtz Instability and Circulation Transfer...]{Kelvin-Helmholtz Instability and Circulation Transfer at an Isotropic-Anisotropic Superfluid Interface in a Neutron Star}
\author[A. Mastrano and A. Melatos]{A. Mastrano$^{1}$\thanks{E-mail:
a.mastrano@physics.unimelb.edu.au} and A.
Melatos$^{1}$\thanks{E-mail: a.melatos@physics.unimelb.edu.au}\\
$^{1}$School of Physics, University of Melbourne, Parkville VIC
3010, Australia}

\begin{document}

\date{Accepted ?. Received ?; in original form ?}

\pagerange{\pageref{firstpage}--\pageref{lastpage}} \pubyear{?}

\maketitle

\label{firstpage}

\begin{abstract}

\noindent{A recent laboratory experiment \citep{4} suggests that a
Kelvin-Helmholtz (KH) instability at the interface between two
superfluids, one rotating and anisotropic, the other stationary
and isotropic, may trigger sudden spin-up of the stationary
superfluid. This result suggests that a KH instability at the
crust-core ($^1S_0$-$^3P_2$-superfluid) boundary of a neutron star
may provide a trigger mechanism for pulsar glitches. We calculate
the dispersion relation of the KH instability involving two
different superfluids including the normal fluid components and
their effects on stability, particularly entropy transport. We
show that an entropy difference between the core and crust
superfluids reduces the threshold differential shear velocity and
threshold crust-core density ratio. We evaluate the wavelength of
maximum growth of the instability for neutron star parameters and
find the resultant circulation transfer to be within the range
observed in pulsar glitches.}

\end{abstract}


\section{Introduction}

It has been proposed that the neutrons making up the inner crust
and core of a neutron star display superfluid-like behaviour
\citep{6,10}. They are compressed to such a high density that they
act like composite bosons, forming Bardeen-Cooper-Schrieffer (BCS)
pairs which condense into the ground state below a critical
temperature $T_c$ and behave like terrestrial superfluids (e.g.
$^4$He). Calculations of the pair interaction and energy gap
suggest that different types of Cooper pairs populate the crust and the
core of a neutron star, $^3P_2$ (triplet) pairs in the core and
$^1S_0$ (singlet) pairs in the crust \citep{25,6,7,9,10,27}.
Superfluidity enhances neutron star cooling, reduces the heat
capacity of the star, and suppresses neutrino emission
\citep{10,17}. In addition, pulsar glitches (discontinuous spin-up
events) are attributed to the motion of superfluid vortex lines;
these lines tend to be pinned to the nuclei in the stellar crust and sudden, large-scale creep of these lines from one
pinning site to another may be responsible for glitches \citep{56}

Macroscopic phase coherence endows a superfluid with frictionless
flow and irrotationality \citep{29,31,32,33}. Irrotationality
leads to formation of discrete quantized vortices if the fluid
rotates. This behaviour is well known and observed in terrestrial
samples of superfluid $^4$He \citep{29,31}. Landau (1941) put
forth what is known as the two-fluid model (where $^4$He is
modelled as two separate components, the inviscid superfluid and
the viscous normal fluid) to explain the various phenomena
observed in superfluid flows such as second sound (an
entropy-temperature wave, where the two components move in
opposite directions, induced by a temperature gradient) and the
fountain effect \citep{29,32}. This hydrodynamic model is
successful and has been polished to a high degree of
sophistication, taking into account more and more effects, such as
the motion of quantized vortices and the drag forces they exert
\citep{41,40b}. Superfluidity in $^3$He, on the other hand, was
first observed more recently \citep{35,36} (see also \citealt{38}
for a first-person account of its discovery) and the search for a
complete hydrodynamic theory continues; for a comprehensive and
up-to-date account, see \citet{33}.

$^3$He superfluid is a condensate of $^3$He Cooper pairs. The
generally accepted theory of $^3$He superfluidity postulates
$p$-wave pairing with spin angular momentum $\hbar$ and orbital
angular momentum $\hbar$. Experiments show that there are two
distinct phases of superfluid $^3$He, known as He-A and He-B. He-B
occupies the $^3P_0$ state; it is isotropic and behaves like
$^4$He superfluid. He-A occupies the $^3P_2$ state; the axis of
the orbital angular momentum defines a preferred direction
(generally denoted as $\bf{\hat{l}}$), breaking the symmetry of
the ground state.\footnote{The experimental identification of the
He-A and He-B as anisotropic and isotropic Cooper pair superfluids
remains ambiguous \citep{36,54}.} This anisotropy manifests itself
as a `texture', defined to be the vector field tangent to the
local direction of the pair angular momentum \citep{35,38,33}. The
texture in the lowest energy state (undisturbed He-A) is uniform,
but heat and fluid flows, applied electromagnetic fields, and
container walls deform it. The texture, in turn, inhibits flows
and heat transport perpendicular to itself. Hence, the presence of
textures modifies the characteristics of any flows induced upon
the fluid.

A recent laboratory experiment with superposed He-A and He-B in
differential rotation (at an interface stabilised by an applied
magnetic field) indicates that the Kelvin-Helmholtz (KH)
instability at the interface plays a significant role in
triggering transfer of circulation from He-A to the initially
irrotational He-B, spinning up the He-B sample; this is detected
as discrete jumps in nuclear magnetic resonance absorption signals
\citep{4,40}. This exciting result suggests that a similar process
may occur at the crust-core superfluid interface in a neutron
star, with important implications for pulsar glitches.

In Section 2 we present an analysis of the KH instability
involving two superposed superfluids, taking as our starting point
the two-fluid model, modified for $^3$He. Our analysis extends
earlier work in two ways: we treat the flow of entropy in the
system self-consistently (cf. \citealt{40}) and consider two
different superposed fluids (each with its own superfluid and
normal fluid components) in relative motion, not the superfluid
and normal fluid components of one fluid \citep{57}. In section 3,
we apply our KH dispersion relation to the experiment conducted by
Blaauwgeers et al. (2003) and to an idealised neutron star model.
This work complements an analysis by \citet{55} of the two-stream
instability, a different process involving two interpenetrating,
isotropic superfluids, which is also a possible triggering
mechanism for glitches.

\section[]{Kelvin-Helmholtz Instability}
\subsection{Two-fluid model of $^3$He}

We restrict our attention to the hydrodynamic regime, where
disturbances of the equilibrium are much slower than the typical
particle collision time. The nondissipative hydrodynamic equations
describing an anisotropic superfluid are (\citealt{29,30,33})

\begin{equation}
\rho_n \left(\frac{\partial {{v}}_{n,i}}{\partial
t}+{\bf{v}}_{n}\cdot\nabla{{v}}_{n,i}\right)= -
\frac{\rho_n}{\rho} \frac{\partial p}{\partial x_i} - s \rho_s
\frac{\partial T}{\partial x_i} - \frac{\rho_s \rho_n}{\rho}
({\bf{v}}_{ns}\cdot\nabla)
{{v}}_{ns,i}+\rho_n{{g}}_i+\sigma_{n,i}\left(\frac{\partial^2
z'_{n}}{\partial x^2}+\frac{\partial^2z'_{n}}{\partial
y^2}\right)\delta(z),\end{equation}
\begin{equation} \rho_s\left(\frac{\partial v_{s,i}}{\partial
t}+{\bf{v}}_s\cdot\nabla v_{s,i}\right) = \frac{\rho_s\hbar}{2m}\epsilon_{jkm}\hat{l}_j\frac{\partial \hat{l}_k}{\partial t}\frac{\partial \hat{l}_m}{\partial x_i}-\frac{\rho_s}{\rho}\frac{\partial p}{\partial
x_i}+s\rho_s\frac{\partial T}{\partial
x_i}+\frac{\rho_s\rho_n}{\rho}{({\bf{v}}_{ns}\cdot\nabla)
v}_{ns,i}+\rho_sg_i +\sigma_{s,i}\left(\frac{\partial^2
z'_{s}}{\partial x^2}+\frac{\partial^2z'_{s}}{\partial
y^2}\right)\delta(z),\end{equation}
\begin{equation}
\left(\frac{\partial}{\partial t} +
{\bf{v}}_n\cdot\nabla\right)\rho s = 0,\end{equation}
\begin{equation}\left(\frac{\partial}{\partial t} +
{\bf{v}}_n\cdot\nabla\right){{\hat{l}}}_i=-\lambda_4\frac{\hbar\rho_s}{2m}\epsilon_{ijk}\hat{l}_j({\bf{v}}_{ns}\cdot\nabla)\hat{l}_k,\end{equation}
\begin{equation}
\nabla\cdot{\bf{v}}_n=0,\end{equation}
\begin{equation}
\nabla\cdot{\bf{v}}_s=0,\end{equation}
\begin{equation}
\left(\frac{\partial}{\partial t} +
{\bf{v}}_n\cdot\nabla\right)\rho_n=0,\end{equation}
\begin{equation} \left(\frac{\partial}{\partial
t}+{\bf{v}}_s\cdot\nabla\right)\rho_s=0,
\end{equation}
\begin{equation} \left(\frac{\partial}{\partial
t}+{\bf{v}}_n|_{z=0}\cdot\nabla\right)z'_n=v'_{n,z}|_{z=0},
\end{equation}
\begin{equation} \left(\frac{\partial}{\partial
t}+{\bf{v}}_s|_{z=0}\cdot\nabla\right)z'_s=v'_{s,z}|_{z=0}.
\end{equation}
The interface is defined to be at $z=0$, the zeroth order flows
are in the $x$-direction, and the velocities in (9) and (10) are
evaluated at the unperturbed interface. In (1)--(10), $\rho_{s,n}$
are the super- and normal fluid densities respectively, $\bf{g}$
is the gravitational acceleration (directed towards negative
$z$-axis), $\bsigma_{n,s}$ is the surface tension force per unit
length for the normal and superfluid interfaces respectively
(directed towards the positive $z$-axis), $z'_{n,s}$ is the
perturbed $z$-coordinate of the normal and superfluid interfaces,
$p$, $T$, $s$, and $m$ denote the pressure, temperature, specific
entropy, and mass of an individual constituent particle,
$\bf{\hat{l}}$ is the unit vector tangent to the texture (in other
words, along the direction of the pair angular momentum),
${\bf{v}}_{s,n}$ are the superfluid and normal fluid velocities,
and ${\bf{v}}_{ns}= {\bf{v}}_s-{\bf{v}}_n$ is the counterflow
velocity. Note that there is no unique equation of state relating
$\rho_s,\rho_n,s,T,$ and $p$; these quantities evolve
independently through the heat transfer (energy) equation (3). An
isotropic superfluid is a special case, in which one sets
${\bf{\hat{l}}}=0$ in (1)--(10).

Equations (1) and (2), the equations of motion governing the
normal fluid and the superfluid, are modified with respect to a
classical fluid by adding the terms $s \rho_s \partial T/\partial
x_i$ and $(\rho_s \rho_n/\rho) ({\bf{v}}_{ns}\cdot\nabla)
{{v}}_{ns,i}$. These terms come from the gradient of the
superfluid chemical potential $\mu$. In a classical fluid, one
simply has $\nabla\mu=\nabla p$, whereas in a superfluid, the
entropy gradient can drive flows separately from the pressure
gradient, because entropy is carried only by the normal fluid
component. The $s \rho_s \partial T/\partial x_i$ term leads to
effects like second sound, a temperature-entropy wave which
differs from `first sound', the usual pressure-density wave
\citep{34}. In addition, the first term on the right-hand side of
(2) describes the force exerted by the texture in an anisotropic
superfluid when the texture is bent; the ground state of the superfluid
far from any walls corresponds to uniform $\bf{\hat{l}}$ (see
Section 2.2), so the texture `stiffens' the system and opposes
flows which might bend it \citep{33}. Equation (3) is our
(non-dissipative) equation of entropy transport [as for the
electron-and-muon fluid in \citet{60}]. Equation (4) describes the
evolution of the texture. Bending the texture (i.e. changing the
direction of $\bf{\hat{l}}$) shifts the energy gap and the momentum distribution of normal fluid excitations, which then reacts back on the
texture itself. The term on the
right-hand side of (4) derives from the dependence of the free energy
density (which is a function of $\partial \hat{l}_i/\partial x_j$) on the superfluid velocity \citep{33}. The remaining
equations are standard: (5) and (6) enforce incompressibility, (7)
and (8) enforce mass continuity, and (9) and (10) ensure that no
fluid particles cross the interface, i.e. the fluids are
immiscible \citep{39}.

For simplicity, we make some approximations. We set all
coefficients of conductivity on the right-hand side of (3) to
zero. We also neglect the anisotropy in density for flows along
and transverse to the texture, and we omit terms involving changes
in free energy as the local texture changes, i.e. we postulate
that the texture is in thermodynamic equilibrium to the zeroth
order. In reality, (2) features a heat current term proportional
to $\nabla T$, with the thermal conductivities for flows parallel
and perpendicular to $\bf{\hat{l}}$ scaling as $\kappa_{\parallel}
\sim (T_c/T) \kappa_N(T_c)$ and $\kappa_{\perp} \sim (T/T_c)
\kappa_N(T_c)$ respectively, where $\kappa_N(T_c)$ is the
(isotropic) conductivity at the critical temperature. Similarly,
although the physical density of the fluid at any point is
uniquely defined, the anisotropic texture modifies the mass
current to become $\rho_\perp v$ perpendicular to $\bf{\hat{l}}$
and $\rho_\parallel v$ parallel to $\bf{\hat{l}}$. This
counterintuitive property, which arises because the two-fluid
description is an effective theory where it is not possible to
label individual $^3$He atoms as {\it{s}} or {\it{n}}, is more
pronounced at lower temperatures, with
$\rho_{n\parallel}=\pi^2(m^*/m)(k_BT/\Delta_0)^2\rho$
and
$\rho_{n\perp}=(7\pi^4/15)(m^*/m)(k_BT/\Delta_0)^4\rho$
as $T \rightarrow 0$, for example. Here, $m^*$ is the effective
mass of each constituent particle and $\Delta_0$ is the maximum
gap parameter, the energy needed to excite one pair out of the
ground state \citep{33}.

\subsection{Unperturbed interface: uniform texture}

We investigate the situation where an isotropic superfluid is
superposed on an anisotropic superfluid and moves relative to it
with a constant and uniform velocity. We follow the derivation of
the dispersion relation of the classical KH instability described
by \citet{39}, assuming continuous stratification of the fluid
variables in the top and bottom layers with a discontinuous jump
at the interface. The normal and superfluid components
interpenetrate; their surfaces are allowed to oscillate with
different amplitudes, but must be in phase with each other.

It is clear that a situation where all the quantities are uniform
and constant in time satisfies (1)--(10). We take this equilibrium
solution as our starting point. A uniform and constant texture,
along with uniform and constant normal fluid velocity, satisfies
(4) identically. A nonuniform but constant texture is also
possible. In fact, it can be shown that the most energetically
favourable texture for a superfluid in a spherical container is
one where the texture lines radiate from a singularity on the
wall, called a `bouquet', `fountain', or `boojum' \citep{33}. In
this paper we assume a uniform and constant texture without any
singularities, as well as uniform and constant fluid velocities.

\subsection{Perturbed interface: surface Kelvin-Helmholtz wave}

We linearise (1)--(10), and take all perturbed (first order)
quantities to be proportional to $\exp(ik_xx+ik_yy-i\omega t)$,
assuming that the wavelength of the perturbation is much smaller
than the length-scale over which the texture changes. We take all
zeroth order quantities (except temperature, which is the same for
both fluids) to be constant and uniform except in the $z$
direction, allowing for vertical stratification. The zeroth-order
shear velocity is chosen in the $x$ direction. From (2), one can
see that, given a spatially uniform zeroth-order texture that is
also constant in time, the textural contribution to the momentum
flux is of quadratic order and thus negligible. One can now ignore
the textural effects on the fluid flows, although the first-order
texture still evolves according to (4). Writing
$D=\partial/\partial z$, we then obtain

\begin{equation}
i\rho^0_n(-\omega+k_xU_n)u_n+\rho^0_n(DU_n)w_n=\frac{-\rho^0_n}{\rho^0_n+\rho^0_s}
(ik_x)p'-s^0\rho^0_s(ik_x)\tau-\frac{\rho_s^0\rho^0_n}{\rho^0_s+\rho_n^0}
[(U_s-U_n)ik_x(u_s-u_n)+(w_s-w_n)D(U_s-U_n)],
\end{equation}
\begin{equation}
i\rho^0_n(-\omega+k_xU_n)v_n=\frac{-\rho^0_n}{\rho^0_n+\rho^0_s}
(ik_y)p'-s^0\rho^0_s(ik_y)\tau-\frac{\rho_s^0\rho^0_n}{\rho^0_n+\rho_s^0}(U_s-U_n)ik_x(v_s-v_n),
\end{equation}
\begin{equation}
i\rho^0_n(-\omega+k_xU_n)w_n=\frac{-\rho^0_n}{\rho^0_n+\rho^0_s}
Dp'-s^0\rho^0_sD\tau-g\rho'_n-\frac{\rho_s^0\rho^0_n}{\rho^0_n+\rho_s^0}(U_s-U_n)ik_x(w_s-w_n)-k^2\sigma_nz'_n\delta(z),
\end{equation}
\begin{equation}
i\rho^0_s(-\omega+k_xU_s)u_s+\rho^0_s(DU_s)w_s=\frac{-\rho^0_s}{\rho^0_n+\rho^0_s}
(ik_x)p'+s^0\rho^0_s(ik_x)\tau+\frac{\rho_s^0\rho^0_n}{\rho^0_s+\rho_n^0}
[(U_s-U_n)ik_x(u_s-u_n)+(w_s-w_n)D(U_s-U_n)],
\end{equation}
\begin{equation}
i\rho^0_s(-\omega+k_xU_s)v_s=\frac{-\rho^0_s}{\rho^0_n+\rho^0_s}
(ik_y)p'+s^0\rho^0_s(ik_y)\tau+\frac{\rho_s^0\rho^0_n}{\rho^0_n+\rho_s^0}(U_s-U_n)ik_x(v_s-v_n),
\end{equation}
\begin{equation}
i\rho^0_s(-\omega+k_xU_s)w_s=\frac{-\rho^0_s}{\rho^0_n+\rho^0_s}
Dp'+s^0\rho^0_sD\tau-g\rho'_s+\frac{\rho_s^0\rho^0_n}{\rho^0_n+\rho_s^0}(U_s-U_n)ik_x(w_s-w_n)-k^2\sigma_sz'_s\delta(z),
\end{equation}
\begin{equation}
(k_xu_{n,s}+k_yv_{n,s})=iDw_{n,s},
\end{equation}
\begin{equation}
i(-\omega+k_xU_{n,s})\rho'_{n,s}=-w_{n,s}D\rho^0_{n,s},
\end{equation}
\begin{equation}
i[-\omega+k_xU_{n,s}(z=0)]z'_{n,s}=w_{n,s}(z=0),
\end{equation}
where $U_{n,s}$ are the zeroth-order shear velocities (in the $x$
direction as defined previously), $u, v,$ and $w$ are the
perturbed velocities in the $x$, $y$, and $z$ directions, $p'$ is
the first order pressure, $\tau$ is the first order temperature,
$\rho'$ is the first order density, and other quantities with the
superscript `0' are of zeroth order. Following \citet{39}, we
eliminate all first-order quantities except for $w_{n,s}$. Using
(11), (12), (14), and (15) we eliminate $\tau$ and $p'$, while
$u_{n,s}$ and $v_{n,s}$ are eliminated using (17). We then obtain
the bulk fluid equations (equations which are valid away from the
interface) from (13) and (16), after using (18) to eliminate
$\rho'_{n,s}$ and (19) to eliminate $z'_{n,s}$.

For the bulk fluids, where all zeroth-order velocities are
constant and uniform everywhere, one obtains the following
equations which are valid throughout the superfluids and normal
fluids away from the interface:

\[\frac{\rho^0_n}{\rho^0}D[\rho^0_s(-\omega+k_xU_s)Dw_s-\rho^0_sk_x(DU_s)w_s]-\frac{\rho^0_s\rho^0_n}{\rho^0}k_x(U_s-U_n)k^2w_s\]
\[\quad +s^0\rho^0_sD\{\frac{\rho^0_n}{s^0\rho^0}[-(-\omega+k_xU_s)Dw_s+k_x(DU_s)w_s+k_x(U_s-U_n)Dw_s-k_x(DU_s-DU_n)w_s]\}\]
\[=\rho^0_n(-\omega+k_xU_n)k^2w_n-\frac{\rho^0_n}{\rho^0}D[\rho^0_n(-\omega+k_xU_n)Dw_n-\rho^0_nk_x(DU_n)w_n]+gk^2(D\rho^0_n)\frac{w_n}{-\omega+k_xU_n}-\frac{\rho^0_s\rho^0_n}{\rho^0}k_x(U_s-U_n)k^2w_n\]
\[\quad -s^0\rho^0_sD\{\frac{\rho^0_n}{s^0\rho^0}[(-\omega+k_xU_n)Dw_n-k_x(DU_n)w_n-k_x(U_s-U_n)Dw_n+k_x(DU_s-DU_n)w_n]\}\]
\begin{equation}\quad -k^4\sigma_n\left(\frac{w_n}{-\omega+k_xU_n}\right)_{z=0}\delta(z)\end{equation}
and

\[\rho^0_s(-\omega+k_xU_s)k^2w_s-\frac{\rho^0_s}{\rho^0}D[\rho^0_s(-\omega+k_xU_s)Dw_s-\rho^0_sk_x(DU_s)w_s]+gk^2(D\rho^0_s)\frac{w_s}{-\omega+k_xU_s}-\frac{\rho^0_s\rho^0_n}{\rho^0}k_x(U_s-U_n)k^2w_s\]
\[\quad +s^0\rho^0_sD\{\frac{\rho^0_n}{s^0\rho^0}[-(-\omega+k_xU_s)Dw_s+k_x(DU_s)w_s+k_x(U_s-U_n)Dw_s-k_x(DU_s-DU_n)w_s]\}\]
\[\quad -k^4\sigma_s\left(\frac{w_s}{-\omega+k_xU_s}\right)_{z=0}\delta(z)\]
\[=\frac{\rho^0_s}{\rho^0}D[\rho^0_n(-\omega+k_xU_n)Dw_n-\rho^0_nk_x(DU_n)w_n]-\frac{\rho^0_s\rho^0_n}{\rho^0}k_x(U_s-U_n)k^2w_n\]
\begin{equation}\quad -s^0\rho^0_sD\{\frac{\rho^0_n}{s^0\rho^0}[(-\omega+k_xU_n)Dw_n-k_x(DU_n)w_n-k_x(U_s-U_n)Dw_n+k_x(DU_s-DU_n)w_n]\}.\end{equation}

From (20) and (21), we see that, away from the interface, where
$U_{n,s}$, $\rho^0_{n,s}$, and $s^0$ are constant and uniform, $w_{n}$ and $w_s$ must satisfy

\begin{equation}
\frac{\rho^0_n\rho^0_s}{\rho^0}k_x(U_s-U_n)(D^2w_s-k^2w_s)=[\frac{\rho^0_n\rho^0_s}{\rho^0}k_x(U_s-U_n)-\rho^0_n(-\omega+k_xU_n)](D^2w_n-k^2w_n),
\end{equation}
and
\begin{equation}
[\frac{\rho^0_n\rho^0_s}{\rho^0}k_x(U_s-U_n)-\rho^0_s(-\omega+k_xU_s)](D^2w_s-k^2w_s)=\frac{\rho^0_n\rho^0_s}{\rho^0}k_x(U_s-U_n)(D^2w_n-k^2w_n).
\end{equation}
One nontrivial, stratified solution is $w_{n,s}$:
$w_{n,B}=N(-\omega+k_xU_{n,B})e^{-kz},
w_{s,B}=S(-\omega+k_xU_{s,B})e^{-kz},
w_{n,A}=N(-\omega+k_xU_{n,A})e^{kz}$, and
$w_{s,A}=S(-\omega+k_xU_{s,A})e^{kz}$, where the fluids on top
(bottom) are labelled with subscript $B$ ($A$). Another solution,
$\omega=k_xU_n\pm(\rho^0_s/\rho^0)^{1/2}k_x(U_n-U_s)$, describes a
wave propagating in the bulk fluid; it does not describe an
instability, so we are not concerned with it here.

The zeroth-order quantities (except $T$, by assumption) jump
across the interface. We integrate the equations across the
interface to analyse the interfacial waves. Many of the terms in
the above equations are of the form $f(z)Dg(z)$, where both $f(z)$
and $g(z)$ are discontinuous functions. Direct integration of such
terms is impossible in general (see Appendix A). However, if one
assumes all the zeroth-order quantities change over the same
length scale $\epsilon$ in passing from the top fluid to the
bottom fluid, we can integrate (20) and (21) term by term from
$z=-\epsilon$ to $z=+\epsilon$ and then squeeze $\epsilon$ to zero
in a mathematically regular way. One then substitutes the
solutions for $(w_{n,s})_{A,B}$ into the integrated (20) and (21)
and obtains two equations involving the oscillation amplitudes $N$
and $S$. The two equations must be compatible, i.e. linearly
dependent. The condition for compatibility gives the following
dispersion relation for waves at the interface:

\[0=[\rho^0_{s,B}(-\omega+k_xU_{s,B})^2+\rho^0_{s,A}(-\omega+k_xU_{s,A})^2+gk(\rho^0_{s,B}-\rho^0_{s,A})-k^3\sigma_s]\]
\[\phantom{++}\times
[\frac{gk}{2}(\rho^0_{s,B}+\rho^0_{s,A})(\rho^0_{n,B}-\rho^0_{n,A})-\frac{k^3}{2}(\rho^0_{s,A}+\rho^0_{s,B})\sigma_n+\alpha\gamma]\]
\[\quad +[\rho^0_{n,B}(-\omega+k_xU_{n,B})^2+\rho^0_{n,A}(-\omega+k_xU_{n,A})^2+gk(\rho^0_{n,B}-\rho^0_{n,A})-k^3\sigma_n]\]
\begin{equation}\phantom{++}\times
[\frac{gk}{2}(\rho^0_{n,B}+\rho^0_{n,A})(\rho^0_{s,B}-\rho^0_{s,A})-\frac{k^3}{2}(\rho^0_{n,A}+\rho^0_{n,B})\sigma_s+\alpha\beta],\end{equation}
with

\[\alpha =
\frac{1}{24}[(s_B+s_A)(\rho^0_B-\rho^0_A)(\rho^0_{s,B}-\rho^0_{s,A})+(s_B-s_A)(\rho^0_B+\rho^0_A)(\rho^0_{s,B}-\rho^0_{s,A})\]
\begin{equation}\phantom{++}+(s_B-s_A)(\rho^0_B-\rho^0_A)(\rho^0_{s,B}+\rho^0_{s,A})]+\frac{1}{8}(s_B+s_A)(\rho^0_B+\rho^0_A)(\rho^0_{s,B}+\rho^0_{s,A}),\end{equation}
\[\beta=
\frac{\rho^0_{n,B}}{s_B\rho^0_B}[(-\omega+k_xU_{s,B})^2-k_x(U_{s,B}-U_{n,B})(-\omega+k_xU_{s,B})]\]
\begin{equation}\phantom{++}+\frac{\rho^0_{n,A}}{s_A\rho^0_A}[(-\omega+k_xU_{s,A})^2-k_x(U_{s,A}-U_{n,A})(-\omega+k_xU_{s,A})],\end{equation}
\[\gamma=
\frac{\rho^0_{n,B}}{s_B\rho^0_B}[(-\omega+k_xU_{n,B})^2-k_x(U_{s,B}-U_{n,B})(-\omega+k_xU_{n,B})]\]
\begin{equation}\phantom{++}+\frac{\rho^0_{n,A}}{s_A\rho^0_A}[(-\omega+k_xU_{n,A})^2-k_x(U_{s,A}-U_{n,A})(-\omega+k_xU_{n,A})],\end{equation}
\[\rho_{A,B}^0=\rho_{s,A,B}^0+\rho_{n,A,B}^0.\]

The dispersion relation (24) is a quartic in $\omega$. It is
related to the dispersion relation of the classical KH instability
as follows. In (24), we have terms involving superfluid and normal
fluid quantities, as well as new terms $\alpha\beta$ and
$\alpha\gamma$ which involve the ratios of the bulk entropies. If
the superfluid and normal fluid interfaces are decoupled, one must
treat the interfaces separately using the classical analysis, i.e.
excluding (3) and entropy terms in (1) and (2). For example, if
either component is negligible, setting either $\rho^0_s$ or
$\rho^0_n$ satisfies (24) identically, or if the interfaces are
allowed to oscillate separately, cross terms like $s\rho_s\partial
T/\partial x$ in (1) or $\rho_s\rho_n{\bf{v}}_{ns}$ in (2) are
absent. If neither component is negligible but the counterflow
tends to zero, setting $\alpha$ to zero in (24) yields the
classical limit, where the components essentially move together as
a classical fluid and entropy does not affect stability. If
neither component is negligible and there is significant
counterflow, one must use the two-fluid dispersion relation (24),
as there is no classical analogue.

\subsection{Instability threshold and entropy effects}

\subsubsection{Classical KH instability threshold}

In order to understand the onset of the superfluid KH instability,
we first review the classical case. The KH instability occurs when
two superimposed layers of fluids with different densities are in
relative motion. When the velocity shear exceed a critical value,
the resulting pressure gradient (from Bernoulli's Law) between the
peaks and troughs of an interfacial wave overcomes the surface
tension and gravity and the mode grows exponentially. In the
classical instability, the interface oscillates at a frequency

\begin{equation}
\omega^{\textrm{c}}=-\frac{k^c(\rho_{A}U_{A}+\rho_{B}U_{B})}{\rho_A+\rho_B}\pm\left[\frac{{gk^c(\rho_{A}-\rho_{B})+(k^c)^3\sigma}}{\rho_{A}+\rho_{B}}-\frac{(k^c)^2\rho_{A}\rho_{B}(U_A-U_B)^2}{(\rho_A+\rho_B)^2}\right]^{1/2}\end{equation}
for wavenumber $k^{\textrm{c}}$ (bearing in mind that the subscripts $A$ and $B$ denote the bottom
and top fluids respectively). Given a particular density and shear
velocity profile, the interface is unstable if
$\textrm{Im}(\omega^{\textrm{c}})>0$, which occurs for wave
numbers between

\begin{equation}
k_{\textrm{min}}^{\textrm{c}}=\frac{\rho_A\rho_B(U_A-U_B)^2}{2\sigma(\rho_A+\rho_B)}-\frac{1}{2\sigma}\left[\frac{\rho_A^2\rho_B^2(U_A-U_B)^4}{(\rho_A+\rho_B)^2}-4\sigma
g(\rho_A-\rho_B)\right]^{1/2}\end{equation} and

\begin{equation}
k_{\textrm{max}}^{\textrm{c}}=\frac{\rho_A\rho_B(U_A-U_B)^2}{2\sigma(\rho_A+\rho_B)}+\frac{1}{2\sigma}\left[\frac{\rho_A^2\rho_B^2(U_A-U_B)^4}{(\rho_A+\rho_B)^2}-4\sigma
g(\rho_A-\rho_B)\right]^{1/2}.\end{equation} On the other hand,
given a particular density profile, the minimum velocity shear
needed to destabilise the interface depends on the densities of
the fluids and surface tension according to \citep{39}

\begin{equation}
(U_{A}-U_{B})^2_{\textrm{min}}=\frac{2(\rho_{A}+\rho_{B})}{\rho_{A}\rho_{B}}[\sigma
g(\rho_{A}-\rho_{B})]^{1/2}.
\end{equation}

\subsubsection{Superfluid KH instability: a simple example}

At first glance, the superfluid dispersion relation (24) has
little in common with the classical dispersion relation (28).
However, the underlying physics is similar, as can be seen by
considering the special case $\rho_{s,A}=\rho_{s,B}=\rho_s$,
$\rho_{n,A}=\rho_{n,B}=\rho_n$, $U_{s,A}=U_{n,A}=0$, and
$U_{s,B}=U_{n,B}=U$, which resembles a classical, single-fluid KH
scenario. In this case, gravity cannot act as a stabilising force,
because the densities of the top and bottom fluids are equal.
However, there may easily exist another stabilising force which does not
depend on the difference of densities (in Section 3.1.1, we discuss
the possible origin of such a force). Thus we replace
$g(\rho_{s,A}-\rho_{s,B})$ with $F_s$ and
$g(\rho_{n,A}-\rho_{n,B})$ with $F_n$. The dispersion relation now
simplifies to

\[0=(2\rho_s\omega^2-2\rho_skU\omega+\rho_sk^2U^2-kF_s-k^3\sigma_s)\]
\[\phantom{++}\times[-\frac{2kF_n}{\rho_n}-\frac{2k^3\sigma_n}{\rho_n}+(2+r+\frac{1}{r})\omega^2-2(1+r)kU\omega+(1+r)k^2U^2]\]
\[\phantom{++}+(2\rho_n\omega^2-2\rho_nkU\omega+\rho_nk^2U^2-kF_n-k^3\sigma_n)\]
\begin{equation}\phantom{++}\times[-\frac{2kF_s}{\rho_s}-\frac{2k^3\sigma_s}{\rho_s}+(2+r+\frac{1}{r})\omega^2-2(1+r)kU\omega+(1+r)k^2U^2],\end{equation}
where $r=s_A/s_B$ is the entropy ratio. We can simplify this
further by investigating the limit where the entropy of the bottom
fluid is much greater than that of the top fluid, i.e.
$r\rightarrow\infty$. Our dispersion relation then becomes

\begin{equation}0=[(2\rho_s\omega^2-2\rho_skU\omega+\rho_sk^2U^2-kF_s-k^3\sigma_s)+(2\rho_n\omega^2-2\rho_nkU\omega+\rho_nk^2U^2-kF_n-k^3\sigma_n)](\omega-kU)^2,\end{equation}
whose roots can be found from $(\omega-kU)^2=0$, which represents
a stable mode, and from

\begin{equation}0=2(\rho_s+\rho_n)\omega^2-2(\rho_s+\rho_n)kU\omega+(\rho_s+\rho_n)k^2U^2-k(F_s+F_n)-k^3(\sigma_s+\sigma_n),\end{equation}
which represents an unstable mode if

\begin{equation}k[2(\sigma_s+\sigma_n)k^2-(\rho_s+\rho_n)U^2k+2(F_s+F_n)]<0.\end{equation}
Therefore, the mode becomes unstable when $U$ exceeds

\begin{equation}U_{\textrm{min}}=\left[\frac{16(\sigma_s+\sigma_n)(F_s+F_n)}{(\rho_s+\rho_n)^2}\right]^{1/4}.\end{equation}
Equation (36) reduces to the classical result
$U_{\textrm{min}}^{\textrm{c}}=(16F\sigma/\rho^2)^{1/4}$ for
$\rho_A=\rho_B$ and is similar in form even for
$\rho_A\neq\rho_B$. Therefore, the underlying physics of the
classical and superfluid KH instabilities are similar, except that, in the latter scenario, the fluids consist of coupled normal and superfluid components, only the normal fluid can transport entropy, and the top and bottom superfluids contain different types of Cooper pairs and
hence have different entropies even at the same temperature. The entropy
difference shifts the critical density and shear velocity as
discussed below and in Figs. 1 and 2. Interestingly, we expect
$r\sim 1$ under a wide range of physical conditions, yet in this
regime (32) does not factorize simply into (33). To recover the
classical-like threshold (36), one needs $r \gg 1$, which is
uncommon in nature.

\subsubsection{Superfluid KH instability: general threshold}

In the superfluid KH instability, the entropy ratio
$r=s_{A}/s_{B}$ shifts the instability threshold. In Fig. 1, we
plot the critical (minimum) density ratio $p=\rho_B/\rho_A$ as $r$
changes for three different values of surface tension
$\sigma/(\rho_AU_A^4g^{-1})=0.22$, 1.1, and 17.7, while the top
fluid is initially stationary. If $p<p_{\textrm{min}}$, the
interface is stable. In Fig. 2, we plot the critical (minimum) $U$
as $r$ changes, for $p=0.1$, $0.5$, and $0.8$. If
$U<U_{\textrm{min}}$, the interface is stable. Tables 1 and 2
summarise the parameters used to construct these figures. From
Figs. 1 and 2, we see that the shift of the instability threshold
away from classical expectations increases as $|s_A-s_B|$
increases. Moreover, as $|s_A-s_B|$ increases, the interface
becomes unstable more easily; interfacial waves are induced for a
wider range of $k$, and stability no longer depends solely on
density ratio or velocity shear. Classically, the KH instability
is driven by a difference in velocity, which causes a difference
in pressure between the top and bottom layers of the interface by
Bernoulli's law. In a superfluid, the temperature gradient
$\nabla\tau$ is also a driving force (or, more generally, the
gradient in chemical potential). Thus, a differential shear flow
at the interface drives a temperature difference between the
layers through (3), destabilising the interface. The magnitude of
this effect is related to $s_A/s_B$, since in the momentum
equations (1) and (2) the temperature gradient force is
proportional to entropy.


In our analysis, we ignore the effects of the container wall. This
is justified in the case of the free $^1S_0$-$^3P_2$ interface in a neutron star.
The friction between the fluids and the container wall also lowers
the instability threshold, as concluded by \citet{40}.

\section{Circulation Transfer}

In this section, we calculate the wavelength and growth rate of
the KH instability for the terrestrial \citet{4} experiment and
neutron star crust-core interface. In both cases, we also
calculate the circulation transferred across the interface, which
is proportional to the number of vortex lines per KH wavelength.

The parameters pertaining to the terrestrial experiment \citep{4}
are summarised in Table 3. The superfluid density is drawn from
the table of total fluid densities given by \citet{29}, assuming
normal fluid densities between $10^{-3}\rho_s$ and $10^{-1}\rho_s$
(the superfluid component dominated in the experiment). The
surface tension at the interface is extrapolated from the
experimental results of \citet{50} and \citet{51}, the latter for
small apertures, which range from $3\times10^{-9}$ to
$6\times10^{-9}$ N m$^{-1}$. Taking $\sigma=10^{-8}$ N m$^{-1}$
and applying eq. (2) from \citet{4}, we extrapolate the
stabilising magnetic force density to be 5 N m$^{-3}$.

The parameters pertaining to the neutron star crust-core interface
are summarised in Table 4. Four cases are considered. In Case A,
we investigate the instability at the crust-core interface of a
typical neutron star, with density and shear velocity profile
taken from \citet{8}, \citet{21}, \citet{10}, \citet{18}, and
\citet{13}. The surface tension is extrapolated from \citet{50},
\citet{51}, and \citet{53}. In Cases B, C, and D, we investigate
how the instability behaves as we vary the density ratio $p$ for
$r=0.1$, $1$, and $10$.

\subsection{\citet{4} experiment}

\subsubsection{Experimental parameters}

In this laboratory experiment, He-A and He-B were superposed in
differential rotation. The interface between He-A and He-B was
magnetically stabilised and the container rotated, inducing
rotation in the normal fluid on both sides of the interface, which
then drags the superfluid component of the A-phase into rotation.
The superfluid component of the B-phase was initially stationary.
Above a critical rotational speed, the B-phase superfluid
experienced sudden jumps in rotation speed, measured as jumps in
the B-phase nuclear magnetic resonance absorption signal. This was
interpreted as the transfer of vortex lines from the A-phase into
the B-phase across the destabilized interface \citep{4}.

The experiment was conducted at $T=0.77 T_c$, $p=29$ bar. The
A-phase normal and superfluid components as well as the B-phase
normal fluid rotated together at 0.0039 m s$^{-1}$, while the
B-phase superfluid was initially stationary. The value we adopt
for the superfluid density, $\rho_s=100$ kg m$^{-3}$, is taken
from the table of total fluid densities given in \citet{29}. The
normal fluid density is lower but an exact figure is not quoted;
we try $\rho_n=0.1$, $1.0$, and $10$ kg m$^{-3}$ (labelled as Case
1, 2, and 3 respectively). At the superfluid A-B interface, the
order parameter of the BCS pair fluid (analogous to magnetisation
in a ferromagnet) changes abruptly, causing surface tension
\citep{50,51}. At $T=0.77 T_c$ and $p=$29 bar, experiments measure
$0.3\leq\sigma_s/(10^{-8} \textrm{N m}^{-1})\leq 0.6$; we adopt
$\sigma_s=10^{-8}$ N m$^{-1}$ for simplicity. The magnetic
restoring force density produced by the magnetic susceptibility
gradient across the interface,
$F=(1/2)(\chi_A-\chi_B)\nabla(H_b^2)\approx 5$ N m$^{-3}$, where
$\chi_A$ and $\chi_B$ are the susceptibilities of the A- and
B-phase respectively and $H_b$ is the magnetic field applied, is
estimated by substituting $\sigma_s$ into the approximate formula
put forward by \citet{4} (on the last page of that paper) to
explain the observed number of vortex lines transferred across the
interface.

Gravity cannot act as the restoring force, because the top and
bottom fluids are of the same densities, so the dispersion
relation (24) needs to be modified to replace gravity with the
magnetic force, viz.

\[0=[\rho^0_{s,B}(-\omega+k_xU_{s,B})^2+\rho^0_{s,A}(-\omega+k_xU_{s,A})^2-k^3\sigma_s-kF_s][-\frac{k}{2}(\rho^0_{s,A}+\rho^0_{s,B})F_n-\frac{k^3}{2}(\rho^0_{s,A}+\rho^0_{s,B})\sigma_n+\alpha\gamma]\]
\begin{equation}\phantom{++} +[\rho^0_{n,B}(-\omega+k_xU_{n,B})^2+\rho^0_{n,A}(-\omega+k_xU_{n,A})^2-k^3\sigma_n-kF_n][-\frac{k}{2}(\rho^0_{n,A}+\rho^0_{n,B}F_s-\frac{k^3}{2}(\rho^0_{n,A}+\rho^0_{n,B})\sigma_s+\alpha\beta].\end{equation}
Our model treats the superfluid and normal fluid components
equivalently; we are free to put the surface tension and magnetic
restoring force into $\sigma_s$ and $F_s$ or $\sigma_n$ and $F_n$.
In reality, of course, there is only one surface tension and one
magnetic restoring force, as experiments cannot distinguish
between superfluid and normal fluid interfaces. However, according
to the interpretation of the experiment tendered by \citet{4}, the
A-phase and B-phase normal fluids move together as one continuous
fluid, so we set $\sigma_n=0$, $F_n=0$, $\sigma_s=10^{-8}$ N
m$^{-1}$, and $F_s=5$ N m$^{-3}$ in the absence of more detailed
measurements. We note that the results are insensitive to this
choice: setting $F_n=F_s=5$ N m$^{-3}$ shifts $k_{\textrm{min}}$
by 0.07 per cent for Case 1, 0.6 per cent for Case 2, and 9 per
cent for Case 3.

\subsubsection{Instability threshold and growth rate}

In the \citet{4} experiment, the normal fluid components are
comoving, so the normal fluid interface is absolutely stable.
Nevertheless, just by being there, the normal fluid significantly
affects the instability threshold due to its ability to transport
entropy via (3). This is an important result. We plot the
threshold wavenumber $k_{\textrm{min}}$ versus the entropy ratio
$r$ in Fig. 3. In all three cases (labelled Case 1, 2, and 3), the
interface becomes unstable at $k_{\textrm{min}}$ different from
the classical value (which is $7.3\times10^3$ m$^{-1}$), particularly at
lower $r$, where $k_{\textrm{min}}$ is much less than the
classical value.\footnote{The classical value needs careful
interpretation in this situation, because the \citet{4} experiment
differs slightly from standard KH instability scenarios. The
superfluid-superfluid interface resembles the classical KH setup,
with the top fluid moving and the bottom stationary, but the
normal fluid components are comoving, so the normal fluid
interface is absolutely stable. Consequently, setting $\alpha=0$
in (24) does not lead to the correct classical (nonentropic)
limit. Instead, one should ignore the normal fluid completely and
apply the classical formula to the superfluid-superfluid interface
\citep{4}.} We see that $k_{\textrm{min}}$ increases with $r$,
jumps sharply at log $r\approx-0.6$, then decreases gently; we
find $1.0\leq k_{\textrm{min}}/k^{\textrm{c}}_{\textrm{min}} \leq
1.1$ between $-0.1\leq \log r \leq 0.1$. The jump occurs at a
slightly lower $r$ in Case 3, where $\rho_n$ is smaller than in
Cases 1 and 2. At low $r$, the $k_{\textrm{min}}$ is lowest (i.e.
the interface is least stable) when $\rho_n$ is highest (Case 3),
indicating that the presence of more normal fluid (whose interface
is KH stable by assumption) helps destabilise the superfluid
interface. However, this trend shifts at log $r\geq0.65$, where
more normal fluid density helps stabilise the interface.
Interestingly, the superfluid interface is stable classically for,
say, $F_s\geq 60$ N m$^{-3}$  but not when entropic effects are
included, when it is unstable for log $r\leq -0.6$ and log $r\geq
1.6$.

Classically, the normal fluid is not expected to influence the
stability of the superfluid interface (in this experiment, where
the normal fluid interface is absolutely stable, the normal fluid
seems naively to act as a passive background). Yet our analysis
shows that it does affect the stability significantly, indicating
that it is driven not only by the shear flow, but also by a `drag
effect' \citep{4} between the superfluid and normal fluid
components in the B-phase, the strength of which depends on the
relative amounts of normal fluid and superfluid present. This
effect comes from the term proportional to $v_{ns}$ in (1) and
(2).

In Fig. 4, we plot the growth rate of the instability [the maximum
of Im($\omega$)] for the three cases above. From Fig. 4, for
$r\sim 1$, we obtain a growth time $\sim 22$ ms, close to the
classical value. This maximum value for Im$(\omega)$ is associated
with cutoff wavenumber $k_{\textrm{cutoff}}=4.71\times 10^4$,
$4.70\times 10^4$, and $4.45\times10^4$ m$^{-1}$ for Cases 1, 2,
and 3 respectively. Note that $k_{\textrm{cutoff}}\sim
6 k_{\textrm{min}}$. The instability in Case 3, where $\rho_n$ is
highest, grows most quickly, indicating that the superfluid
interface is destabilised even further by the normal fluid,
through the drag effect. We anticipate that our prediction of the
KH growth time will be tested by future nuclear magnetic resonance
measurements.

\subsubsection{Angular velocity jump}

When the interface becomes unstable, it is hypothesised that superfluid vortex lines break through the interface and transfer circulation from one fluid to the other. As circulation is quantized, the amount
transferred depends on the wavelength of the KH instability, that is,
on how many vortex lines $(\delta N)$ can fit within one corrugation of the wave.
A-phase vortex lines are $\sim 10^3$ times thicker radially than B-phase ones, so a concentration of energy
is needed to push and convert A-phase lines into B-phase lines. This free energy
is provided by the instability at the interface \citep{4}. The
precise mechanism of vortex transfer and conversion is still
unknown, but \citet{40} suggested that the vortices in the A-phase
are pushed by the Magnus force towards the vortex-free
B-phase, entering the potential well formed by the
corrugation of the unstable interface and enhancing the growth of
the well until it forms a droplet of A-phase fluid filled with
vorticity, which then resides in the B-phase \citep{40}.

For Cases 1, 2, and 3, $k_{\textrm{min}}$ at $r\approx 1$ is
approximately equal to the classical value of $7300$ m$^{-1}$,
giving an instability wavelength
$\lambda_0=2\pi/k_{\textrm{min}}=8.6\times 10^{-4}$ m. The amount
of circulation transferred across the interface is
$\kappa=|U_A-U_B|\lambda_0/2$ \citep{40}. The circulation quantum
of superfluid $^3$He is $\kappa_0=h/2m_3=6.6\times10^{-8}$ m$^2$
s$^{-1}$, where $m_3$ is the mass of a $^3$He atom. Our analysis
of the \citet{4} experiment thus implies $\kappa=1.7\times10^{-6}$
m$^2$ s$^{-1}$ and $\delta N = \kappa/\kappa_0 \approx 25$.
Experimentally, \citet{4} observed $\delta N\approx 10$. The
discrepancy may be due, in part, to misestimating of
$\sigma_{n,s}$, $F_{n,s}$, and/or $\rho_{n,s}$ in the absence of
exact quoted values. We take heart, however, from semiquantitative
agreement between theory and experiment and proceed to apply the
theory to the crust-core interface in a neutron star.

\subsection{Neutron star $^1S_0$-$^3P_2$ interface}

\subsubsection{Stellar parameters}

We assume that the phase transition from $^1S_0$ to $^3P_2$
neutron superfluid in a neutron star occurs at a sharp boundary
near the crust-core interface at radius $R\approx 6$ km
\citep{10}. The inner crust consists entirely of neutron BCS pairs
in the $^1S_0$ state, behaving as an isotropic superfluid of
density $\rho_B=1.5 \times 10^{17}$ kg m$^{-3}$
\citep{8,21,10,13}. The temperature of the crust is around $10^8$
K, which is about $0.01 T_c$ \citep{18,10} so the normal fluid is
negligible; we therefore assume\footnote{We attempted to calculate
the exact temperature-dependent value of $\rho_n/\rho_s$ from eqs.
(3.92) and (3.94) in \citet{33}. However, due to uncertainties in
$m^*$ and the Fermi-liquid corrections in a neutron superfluid, we
obtained $\rho_n\approx 0.3\rho_{\textrm{total}}$ at $T=$0.01
$T_c$, which seems to be higher than existing phenomenological
models can sustain \citep{56}. Thus we take
$\rho_n=0.01\rho_{\textrm{total}}$ for now.} $\rho_s=1.5 \times
10^{17}$ kg m$^{-3}$ and $\rho_n=1.5 \times10^{15}$ kg m$^{-3}$ in
the absence of detailed knowledge. The core consists of neutron
BCS pairs in the $^3P_2$ state, with $T=0.2 T_c=10^8$ K
\citep{10}. Hence the superfluid component is dominant here also;
we assume $\rho_s=2.8 \times 10^{17}$ kg m$^{-3}$ and $\rho_n=2.8
\times10^{15}$ kg m$^{-3}$. This density and shear velocity
profile we label as Case A, summarised in Table 4. We also
calculate how the instability wave number $k_{\textrm{min}}$
changes with respect to density ratio
$p=\rho_{\textrm{crust}}/\rho_{\textrm{core}}$ with density ratio
$r$ fixed at 0.1, 1, and 10. We present these results as Cases B,
C, and D respectively in Table 4. Note that, in the subsonic
regime we investigate, the neutron star matter is approximately
incompressible.

The surface tension at the neutron star crust-core interface is
unknown. However, according to \citet{53}, surface tension scales
as $m^* p_{\textrm{F}} T_c^2(1-T/T_c)^{3/2}$, where $m^*$ is the
effective mass of the particles and $p_{\textrm{F}}$ is the Fermi
momentum. Noting that $p_{\textrm{F}}$ in a neutron superfluid is
$\sim 2\times10^5$ times higher than terrestrial $^3$He and $T_c$
is $\sim 10^{11}$ times higher, and terrestrial $^3$He is measured
to have $\sigma_s=10^{-8}$ N m$^{-1}$ \citep{50,51}, we estimate
that the surface tension of neutron star superfluid is
$\sigma_s\sim 10^{19}$ N m$^{-1}$, i.e. 25 orders of magnitude
higher proportionately. We are acutely aware of the shortcomings
of this estimate and hope that more reliable calculations will be
undertaken in the literature in the future.

\subsubsection{Glitch threshold and growth rate}

We plot $k_{\textrm{min}}/k_{\textrm{min}}^\textrm{c}$ versus log
$r$ for Case A in Fig. 5, where
$k_{\textrm{min}}^\textrm{c}\approx 0.67$ m$^{-1}$ is the
classical value of the instability threshold, obtained by setting
$\alpha=0$ in (24). We see that as $r$ increases, the interface
becomes less stable. The instability wave number
$k_{\textrm{min}}$ is inversely proportional to shear velocity $U$
and directly proportional to gravitational acceleration $g$, as
purely classical KH instability theory predicts \citep{39}. At low
$r$, the plot is flat and $k_{\textrm{min}}$ is fairly close to
the classical value ($k_{\textrm{min}}= 2.662$, 0.666, and 0.167
m$^{-1}$ for $U=5 \times 10^{5}$, $10^6$, and $2\times10^6$ m
s$^{-1}$ respectively). Case A is classically unstable, but the
shift in $k_{\textrm{min}}$ is only noticeable at higher $r$,
where the plot dips at log $r\geq -0.5$. Note that we find
$k_{\textrm{min}} R_* \gg 1$, so the assumption of a plane wave
perturbation is justified {\it{a posteriori}}.

One can also vary the density ratio $p$ while keeping $r$
constant. In Fig. 6, we plot $k_{\textrm{min}}$ against $p$ for
three different entropy ratios $r=0.1$, 1, and 10, labelled as
Cases B, C, and D respectively (see Table 4). The classical
minimum wave number approaches infinity as the top fluid density
reaches zero, i.e. the interface becomes absolutely stable as the
top fluid disappears. When entropy is taken into account,
$k_{\textrm{min}}$ still increases as $p$ gets lower, but at a
slower rate, especially at higher $r$. Cases B, C, and D join the
classical curve at $p\approx 0.45$, 0.9, and 1.0, demonstrating
again that we diverge further from the classical expectation as
$|s_A-s_B|$ increases.

In the foregoing, we do not consider the effects of a magnetic
field and susceptibility discontinuity at the interface. In
principle, a magnetic field can stabilise the boundary as in the
Blaauwgeers experiment (see Section 3.1). We can set an
approximate upper bound on $\nabla(H_b^2)$ at $B^2/(\mu^2_0R)$,
where $B$ is the magnetic field at the surface of the star
(typically $10^8-10^9$ T). Thus, the upper bound on the magnetic
force density is between $1.06 \times 10^{24}$ and $1.06 \times
10^{26}$ N m$^{-3}$. However, we find that, for Case A (i.e.
typical neutron star parameters), the minimum magnetic force
density necessary to stabilise the interface occupies the range
$2.4$--$2.9\times10^{38}$ N m$^{-3}$ for $-2\leq \log r\leq 1$,
much higher than the maximum force density available.

For a given shear velocity and density profile, one can now
estimate the growth rate $\gamma$ of the instability. We present
our plot of growth rate versus $r$ for a standard neutron star
density profile (Case A) as Fig. 7; the classical growth rate is
Im$(\omega^{\textrm{c}})=1.77 \times 10^{15}$ s$^{-1}$ for these
parameters. The growth rate is found to be directly proportional
to the cube of the shear velocity and inversely proportional to
surface tension. We find $\gamma\approx 1.01 \gamma^c$ for $r\sim
1$. The growth time grows monotonically with $k$ and takes the
value of $5.59\times10^{-16}$ s at
$k_{\textrm{cutoff}}=6.53\times10^9$ m$^{-1}$ (where
$k_\textrm{cutoff}$ is determined by the surface tension).

If there is no surface tension, the viscosity of the fluid
suppresses the KH instability when the viscous time-scale is
shorter that the growth time of the KH oscillation, i.e. for
$k^2\nu^{-1}\textrm{Im}(\omega)$, where $\nu$ is the kinematic
viscosity. The viscosity of superfluid neutron star matter is
estimated to lie between 0.1 and 0.17 m$^2$ s$^{-1}$ for Case A
\citep{52}. In Fig. 8, we present a plot of Im$(\omega)$ versus
$r$. We find Im$(\omega) \propto U^2 \nu^{-1}$. For Case A, if the
surface tension vanishes but viscosity suppresses the instability,
we estimate Im$(\omega)=1.1 \textrm{Im}(\omega^{\textrm{c}})$ and
$k_{\textrm{cutoff}}=5.24\times10^6$ m$^{-1}$ for $r=1$,
translating into a growth time of $4\times10^{-13}$ s. The growth
time is $\sim 10^3$ times larger than the limit imposed by surface
tension above, indicating that viscosity suppresses the
instability more effectively than surface tension.

{\it{Prima facie}}, our KH growth time estimate seems too small;
certainly, it is unresolvable by radio pulsar timing. However, it
is highly dependent on $k_{\textrm{cutoff}}$
($k_{\textrm{cutoff}}\gg k_{\textrm{min}}$, unlike in the
terrestrial experiment). As $k_\textrm{cutoff}$ depends on
dissipation physics which is largely neglected in our model, e.g.
textural terms on the right-hand sides of (1) and (2), a more
complete analysis is needed before the KH growth time can be
predicted with confidence.

\subsubsection{Angular velocity jump}

For typical neutron star parameters (Case A), near $r=1$, we
obtain $k_{\textrm{min}}\approx 0.587$ m$^{-1}$, and hence
$\lambda_0=10.7$ m. This implies that the amount of circulation
transferred across the interface is
$\kappa=|U_A-U_B|\lambda_0/2=5.35\times10^{6}$ m$^2$ s$^{-1}$. The
total circulation at the interface is roughly $2\pi
R|U_A-U_B|\approx 6\times 10^9$ m$^2$ s$^{-1}$. Therefore, the
fractional change in angular velocity is $\Delta \Omega/\Omega
\approx \lambda_0 R \approx 1.4\times 10^{-4}$. This compares
surprisingly well with observations of the angular velocity jump
in pulsar glitches, viz. $10^{-9}<\Delta \Omega/\Omega<10^{-4}$
\citep{78}.

\subsubsection{Proton superfluid}

Hydrodynamic models of neutron star superfluids in the literature
\citep{60,55} often include protons, electrons, and muons.
The latter two species can be effectively incorporated into the
normal fluid in our analysis \citep{60}, but the protons cannot.
How significant is the proton superfluid?

One key effect of the protons is the drag they exert on the neutrons.
The strength of the `drag effect' between the protons and neutrons is set by the off-diagonal elements of the symmetrical
mass density matrix, i.e.
$\rho_{np}=\rho_{pn}=\rho_n(m_n^*-m_n)/m_n^*=\rho_p(m_p^*-m_p)/m_p^*$,
where $\rho_n$ ($\rho_p$), $m_n$ ($m_p$), and $m^*_n$ ($m^*_p$)
are the neutron (proton) density, bare mass, and effective mass
respectively. Estimates of $m^*_p$ vary between $0.3 m_p$ and $0.7
m_p$, while $\rho_p$ can be estimated from neutron star cooling
rates \citep{80}. If we take the proton density to be 15 per cent
of the total core density of $2.8 \times 10^{17}$ kg m$^{-3}$, the
threshold proton fraction for direct URCA cooling \citep{80}, we
obtain $|\rho_{np}/\rho_n| \sim 0.15$. Hence the proton-neutron
drag force in Eq. (14) of \citet{60}, as a fraction of a typical
term in our momentum equation (1), is given by
$(\rho_{np}/\rho)|{\bf{v}}^{(p)}-{\bf{v}}^{(n)}||\nabla \times
{\bf{v}}^{(n)}|/|{\bf{v}}^{(n)}| \omega \lesssim
(\rho_{np}/\rho)(2kv^{(n)}/\omega)\sim 0.3$ (for a typical KH
instability, the phase velocity is of the same order as the shear
velocity), where ${\bf{v}}^{(n)}$ (${\bf{v}}^{(p)}$) is the
neutron (proton) superfluid velocity. In other words, the
proton-neutron drag is smaller than, but comparable to, the forces
we consider and should be included in general. It is omitted in
this paper to keep the algebra manageable and to maintain focus on the
entropic KH effects, presented here for the first time. Note
that, although the normal fluid fraction under neutron star
conditions may be low, the entropy ratio $r$ is what matters in
our analysis, and it is not negligible.

There are other physical effects that also appear when charged
species are included. In this paper, we intentionally neglect
electromagnetic forces [see Eqs. (15) and (16) in \citet{60}] and
electromagnetic contributions to the chemical potential [Eqs. (29)
and (30) in \citet{60}], again to render the superfluid KH
analysis tractable. One electromagnetic effect is included: the
force term $F_{n,s}$, which can stabilise the interface in
principle (see Section 3.1.1 above), although in practice it is
too weak to shift $k_{\textrm{min}}$
appreciably in a neutron star (see Section 3.2.2). Apart from
excluding the proton superfluid, our hydrodynamic equations are
similar to those written down by \citet{60}. In our case, the
coupling is between the normal fluid and superfluid components and is governed by entropy (this coupling has been unambiguously observed
in laboratory experiments, e.g. second sound in He II). Our terms
involving entropy and counterflow are derived from the gradient of
the chemical potential (see Section 2.1) and therefore are
implicitly included in the neutron superfluid chemical potential
$\tilde{\mu}_n$ [Eq. (29) in \citet{60}]. Another important
difference is that \citet{60} and \citet{55} deal with a
non-stratified neutron star which consists of two different but interpenetrating superfluids, whereas we deal with a stratified (two-zone) neutron star
with two layers of superfluid in differential rotation.

\section{Conclusions}

We analyse the superfluid KH instability within the framework of
two-fluid hydrodynamics including entropy flow and both the
superfluid and normal fluid components in each fluid. We find that
the coupling between the superfluid and the normal fluid
components via entropy transport lowers the instability threshold
(especially when the entropy ratio is large) even if there is no
significant counterflow. When the counterflow is significant, as
in  the experiment conducted by \citet{4}, the reduction in the
threshold is even greater. Classically, the KH instability occurs
because the velocity shear induces a self-reinforcing pressure
gradient across the interface. In a superfluid, the (perturbed)
temperature gradient, driven by the differential flow, is also a
driving force, the magnitude of which is controlled by ratio of
the entropies of the fluids, and the interface becomes less stable
than expected classically.

When the interface crinkles, it is believed that superfluid vortex
lines break through and transfer circulation between the fluids,
although the mechanism is not understood at the Gross-Pitaevskii
level. Since circulation is quantized, the amount transferred
depends on how many vortex lines can fit in one corrugation of the
KH surface wave. Applied to \citet{4} experiment, our model yields
a growth time of $\sim 22$ ms and threshold wavenumber
$k_{\textrm{min}}=7.3\times10^{3}$ m$^{-1}$, so that the number of vortex
lines transferred is $\delta N\approx 25$. The observed number of
vortex lines is 10 \citep{4}. Applied to the neutron star
crust-core ($^1S_0$-$^3P_2$) interface, with $U=10^6$ m sec$^{-1}$
and $r=1$, we find $\lambda_0=11$ m, so that the fractional change
in angular velocity is $\Delta \Omega/\Omega \approx \lambda_0 R
\approx 1.4\times 10^{-4}$. This compares favourably with typical
pulsar glitch events, which are observed to have $10^{-9}<\Delta
\Omega/\Omega < 10^{-4}$ \citep{78}. In view of this estimate, and
the high KH growth rate, it is possible that vortex line transfer
through a surface KH instability may explain pulsar glitches. An
alternative process, the two-stream instability of isotropic
neutron and proton superfluids in differential rotation, is also
an attractive possibility \citep{55}. A more thorough analysis of
both, taking into account the nonuniform texture and anisotropic
densities and thermal conductivities of the core superfluid, is
worth pursuing.

\section*{Acknowledgments}

We thank Derek Chan, Barry Hughes, James McCaw, Bruce McKellar, Andy Martin, Thanu Padmanabhan, and Chris Pakes for discussions.

\begin{table*}
 \centering
 \begin{minipage}{170mm}
  \caption{Fluid parameters for the calculation of threshold density ratio $p=\rho_B/\rho_A$ as a function of entropy ratio $r=s_A/s_B$ in Fig. 1. The top fluid $B$ is stationary. The superfluid and normal fluid components of the bottom fluid $A$ move at the same velocity $U_A$. We investigate three values of $U_A$ (Cases I--III). The classical thresholds are $p^c_{\textrm{min}}=0.997$, $0.947$, and $0.780$ for Cases I- III respectively.}
  \begin{tabular}{@{}lccc@{}}
  \hline
   Parameters&  Case I&Case II&Case III  \\

 \hline
 $\rho_{n,A}/\rho_A$ & $9.90\times10^{-3}$& $9.90\times10^{-3}$&$9.90\times10^{-3}$ \\
 $\rho_{s,A}/\rho_A$ & $9.90\times10^{-1}$&$9.90\times10^{-1}$&$9.90\times10^{-1}$ \\
 $U_{B}/U_A$ & 0&0&0\\
 $\sigma_n[g^{-1}\rho_A U_A^{4}]^{-1}$ & $17.7$&$1.10$&$2.18\times10^{-1}$  \\
 $\sigma_s[g^{-1}\rho_A U_A^{4}]^{-1}$ &$17.7$&$1.10$&$2.18\times10^{-1}$\\

 \hline
\end{tabular}
\end{minipage}
\end{table*}

\begin{table*}
 \centering
 \begin{minipage}{170mm}
  \caption{Fluid parameters for the calculation of threshold velocity shear $U_\textrm{min}$ as a function of entropy ratio $r=s_A/s_B$ in Fig. 1. The top fluid $B$ is stationary. The superfluid and normal fluid components of the bottom fluid $A$ move at the same velocity $U_\textrm{min}$. We investigate three values of $p=0.1$, 0.5, and 0.8 (Cases IV--VI).}
  \begin{tabular}{@{}lccc@{}}
  \hline
   Parameters& Case IV&Case V&Case VI  \\

 \hline
 $\rho_{n,B}/\rho_A$ & $9.90\times 10^{-4}$ &  $4.95\times 10^{-3}$&$7.92\times 10^{-3}$ \\
 $\rho_{s,B}/\rho_A$ & $9.90\times 10^{-2}$& $4.95\times 10^{-1}$&$7.92\times 10^{-1}$\\
 $\rho_{n,A}/\rho_A$ & $9.90\times10^{-3}$& $9.90\times10^{-3}$&$9.90\times10^{-3}$ \\
 $\rho_{s,A}/\rho_A$ & $9.90\times10^{-1}$&$9.90\times10^{-1}$&$9.90\times10^{-1}$ \\
 $\sigma_n[g^{-1}\rho_A (U_{\textrm{min}}^\textrm{c})^{4}]^{-1} $ & $2.29\times10^{-3}$&$5.56\times10^{-2}$ &$2.47\times10^{-1}$ \\
 $\sigma_s[g^{-1}\rho_A (U_{\textrm{min}}^\textrm{c})^{4}]^{-1}$ & $2.29\times10^{-3}$&$5.56\times10^{-2}$ &$2.47\times10^{-1}$\\
 $U_{n,B}/U^\textrm{c}_{\textrm{min}}$ & 0&0&0\\
 $U_{s,B}/U^\textrm{c}_{\textrm{min}}$&0&0&0\\

 \hline
\end{tabular}
\end{minipage}
\end{table*}




\begin{table*}
 \centering
 \begin{minipage}{170mm}
  \caption{Fluid parameters for the instability threshold calculations as a function of entropy ratio $r$ in Figs. 3 and 4. The parameters are taken from the spin-up experiment of \citet{4}. The bottom and top fluids are labelled $A$ and $B$ respectively. We investigate three values of density ratio $p$ (Cases 1--3).}
  \begin{tabular}{@{}lcccc@{}}
  \hline
   Parameters   & Units & Case 1& Case 2& Case 3 \\

 \hline
 $\rho_{n,B}$ &kg m$^{-3}$& 0.1 & 1 & 10   \\
 $\rho_{s,B}$ &kg m$^{-3}$& 100 & 100 & 100 \\
 $\rho_{n,A}$ &kg m$^{-3}$& 0.1 & 1 & 10  \\
 $\rho_{s,A}$ &kg m$^{-3}$ & 100 & 100 & 100\\
 $U_{n,B}$ &m s$^{-1}$& $3.9\times10^{-3}$& $3.9\times10^{-3}$ & $3.9\times10^{-3}$\\
 $U_{s,B}$ &m s$^{-1}$& $3.9\times10^{-3}$& $3.9\times10^{-3}$ & $3.9\times10^{-3}$\\
 $U_{n,A}$ &m s$^{-1}$& $3.9\times10^{-3}$ & $3.9\times10^{-3}$ & $3.9\times10^{-3}$  \\
 $U_{s,A}$ &m s$^{-1}$& 0 & 0 & 0 \\
 $\sigma_n$\footnote{Surface tensions are extrapolated from the experimental results of \citet{50} and \citet{51}.} &N m$^{-1}$& 0 & 0 & 0  \\
 $\sigma_s$ &N m$^{-1}$& $1\times10^{-8}$ & $1\times10^{-8}$ & $1\times10^{-8}$  \\
 $F_n$ \footnote{Magnetic restoring force is inferred from \citet{4} and the surface tension.} &N m$^{-3}$& 0 & 0 & 0  \\
 $F_s$ &N m$^{-3}$& 5 & 5 & 5  \\
 $k_{\textrm{min}}^{\textrm{c}}$ \footnote{Classical minimum instability wavenumber.}&m$^{-1}$& $7.3\times10^{3}$ & $7.3\times10^{3}$ & $7.3\times10^{3}$ \\

\hline
\end{tabular}
\end{minipage}
\end{table*}

\begin{table*}
 \centering
 \begin{minipage}{170mm}
  \caption{Fluid parameters for the instability threshold calculations as a function of entropy ratio $r$ (Case A) and density ratio $p$ (Cases B--D). The crust fluid $(B)$ is initially stationary in all cases; the core $(A)$. In Case A, we investigate how the threshold wavenumber $k_{\textrm{min}}$ changes with entropy ratio $r=s_A/s_B$. In Cases B, C, and D, we investigate how $k_{\textrm{min}}$ changes with density ratio $p=\rho_B/\rho_A$ for $r=0.1$, 1.0, and 10 while keeping the density and velocity of the core constant.}
  \begin{tabular}{@{}lccccc@{}}
  \hline
   Parameters   &Units  & Case A&Case B&Case C&Case D \\

 \hline
 $g$ &m s$^{-2}$ & $5\times10^{11}$ & $5\times10^{11}$ & $5\times10^{11}$&$5\times10^{11}$\\
 $\rho_{n,B}$ &kg m$^{-3}$& $1.5\times 10^{15}$ & $2.8\times10^{14}$--$2.8\times10^{15}$& $2.8\times10^{14}$--$2.8\times10^{15}$&$2.8\times10^{14}$--$2.8\times10^{15}$ \\
 $\rho_{s,B}$ &kg m$^{-3}$&  $1.5\times 10^{17}$& $2.8\times10^{16}$--$2.8\times10^{17}$&$2.8\times10^{16}$--$2.8\times10^{17}$&$2.8\times10^{16}$--$2.8\times10^{17}$\\
 $\rho_{n,A}$ &kg m$^{-3}$& $2.8\times10^{15}$ & $2.8\times10^{15}$ &$2.8\times10^{15}$ &$2.8\times10^{15}$ \\
 $\rho_{s,A}$ &kg m$^{-3}$ &  $2.8\times10^{17}$& $2.8\times10^{17}$ & $2.8\times10^{17}$&$2.8\times10^{17}$\\
 $U_{n,B}$ &m s$^{-1}$&  0 & 0 &0 &0\\
 $U_{s,B}$ &m s$^{-1}$&  0 & 0 & 0&0\\
 $U_{n,A}$ &m s$^{-1}$&  $10^6$ & $10^6$ &$10^6$ &$10^6$ \\
 $U_{s,A}$ &m s$^{-1}$& $10^6$ & $10^6$ & $10^6$& $10^6$\\
 $\sigma_n$\footnote{Surface tensions are extrapolated from \citet{50}, \citet{51}, and the theory in \citet{53}.} &N m$^{-1}$& $10^{19}$ & $10^{19}$&$10^{19}$ & $10^{19}$ \\
 $\sigma_s$ &N m$^{-1}$&  $10^{19}$ & $10^{19}$& $10^{19}$& $10^{19}$ \\
 $F_n$  &N m$^{-3}$&  0 & 0& 0& 0\\
 $F_s$ &N m$^{-3}$&  0 & 0& 0&0 \\
 $r$&---&0.01--10&0.1&1&10\\

\hline
\end{tabular}
\end{minipage}
\end{table*}

\begin{figure}
 \includegraphics{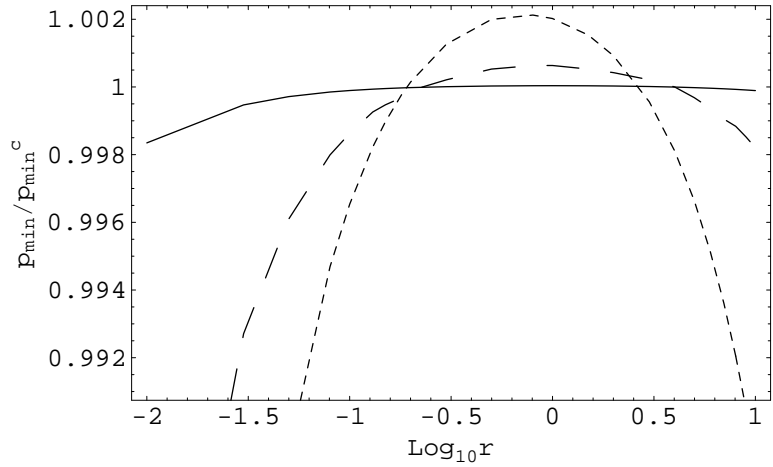}
 \caption{Instability threshold. Minimum density ratios $p_{\textrm{min}}$ (normalised to the classical value $p_{\textrm{min}}^{\textrm{c}}$) plotted against log $r$ for three different shear velocities, as in Table 1: Case I (solid curve, lowest $U$, $p_{\textrm{min}}^{\textrm{c}}=0.997$), Case II (dashed curve, intermediate $U$, $p_{\textrm{min}}^{\textrm{c}}=0.947$), and Case III (dotted curve, greatest $U$, $p_{\textrm{min}}^{\textrm{c}}=0.780$). The interface is stable if the system lies {\it{below}} the curves.}
 \label{fig1}
\end{figure}

\begin{figure}
 \includegraphics{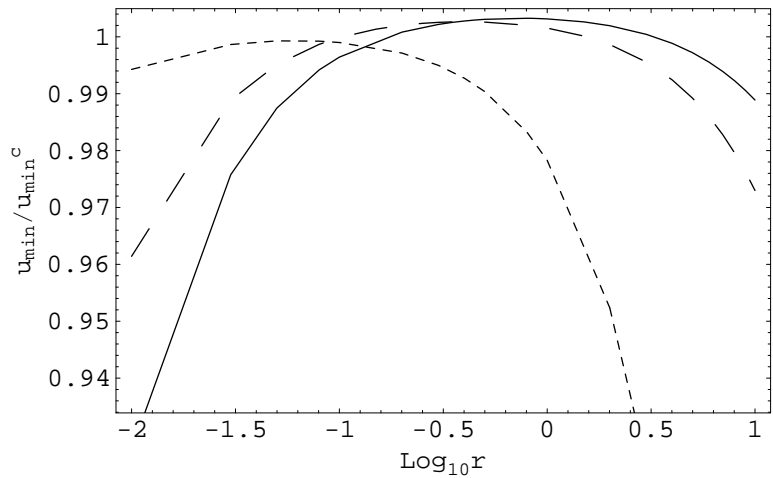}
 \caption{Instability threshold. Minimum shear velocity $U_{\textrm{min}}$ (normalised to the classical value $U_{\textrm{min}}^{\textrm{c}}$ plotted against log $r$ for three different density ratios $p$, as in Table 2: Case VI (solid curve, $p=0.1$), Case V (dashed curve, $p=0.5$), and Case IV (dotted curve, $p=0.8$). The interface is stable if the system lies {\it{below}} the curves}
 \label{fig2}
\end{figure}


\begin{figure}
 \includegraphics{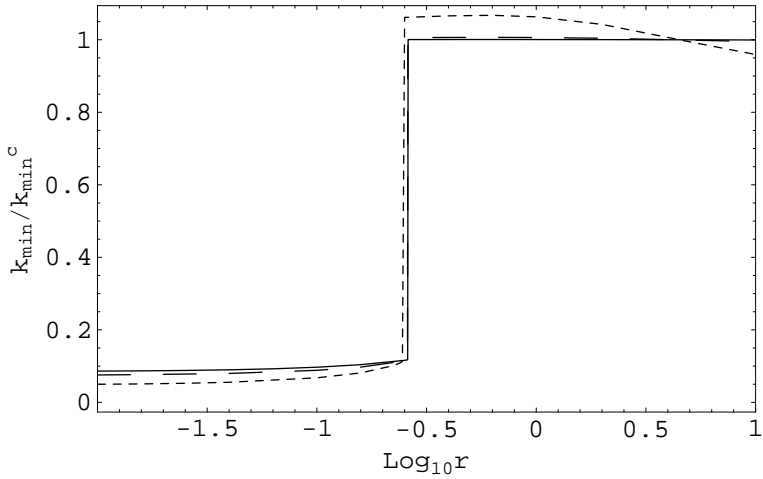}
 \caption{Threshold wavenumber $k_{\textrm{min}}$ versus entropy ratio $r$ for the \citet{4} experiment. The classical threshold wavenumber $k_{\textrm{min}}^{\textrm{c}}=7.3\times10^3$ m$^{-1}$. The solid curve is Case 1 ($\rho_{n,A}=\rho_{n,B}=0.1$ kg m$^{-3}$), the dashed curve is Case 2 ($\rho_{n,A}=\rho_{n,B}=1$ kg m$^{-3}$), and the dotted curve is Case 3 ($\rho_{n,A}=\rho_{n,B}=10$ kg m$^{-3}$).}
 \label{fig3}
\end{figure}





\begin{figure}
 \includegraphics{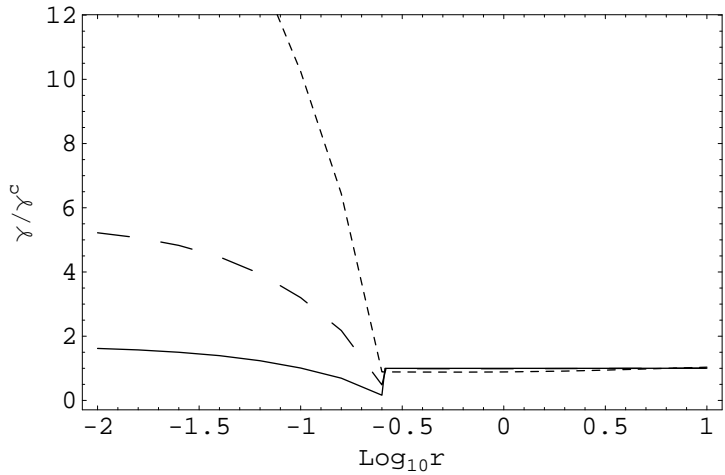}
 \caption{Maximum growth rate $\gamma$ versus entropy ratio $r$ for the \citet{4} experiment, normalised to the classical growth rate $\gamma^{\textrm{c}}=45.10$ s$^{-1}$. The solid curve is Case 1 ($\rho_{n,A}=\rho_{n,B}=0.1$ kg m$^{-3}$), the dashed curve is Case 2 ($\rho_{n,A}=\rho_{n,B}=1$ kg m$^{-3}$), and the dotted curve is Case 3 ($\rho_{n,A}=\rho_{n,B}=10$ kg m$^{-3}$).}
 \label{fig4}
\end{figure}

\begin{figure}
 \includegraphics{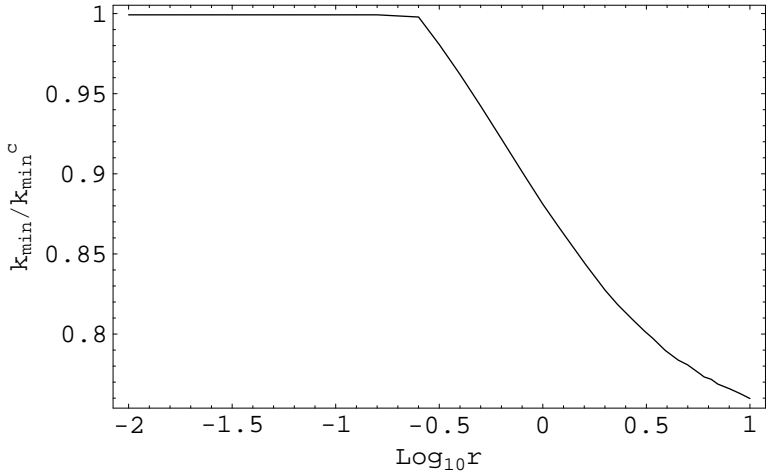}
 \caption{Threshold wavenumber $k_{\textrm{min}}$ versus entropy ratio $r$ at the crust-core boundary of a standard neutron star. This is labelled as Case A ($U_{\textrm{core}}=10^6$ m s$^{-1}$) is Table 4. The classical threshold wavenumber $k_{\textrm{min}}^c=0.67$ m$^{-1}$. We find that $k_{\textrm{min}}
\propto  g U^{-2}$.}
 \label{fig5}
\end{figure}

\begin{figure}
 \includegraphics{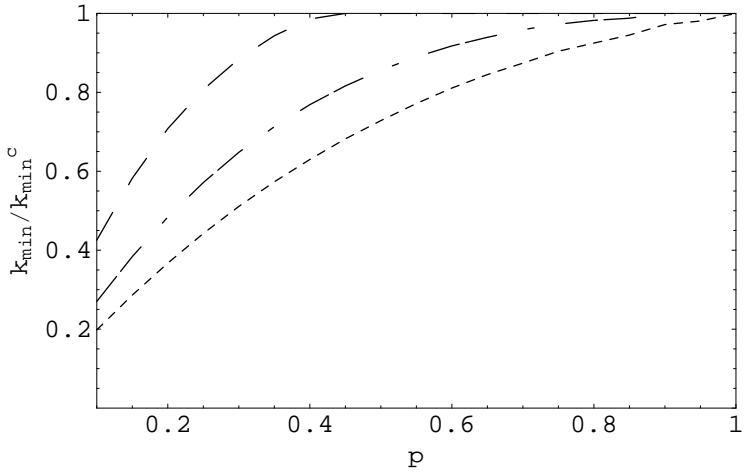}
 \caption{Threshold wavenumber $k_{\textrm{min}}$ versus density ratio $p$ at the crust-core boundary of a neutron star. The dashed curve is Case B ($r=0.1$), the dashed-dotted curve is Case C ($r=1$), and the dotted curve is Case
 D ($r=10$).}
 \label{fig6}
\end{figure}


\begin{figure}
 \includegraphics{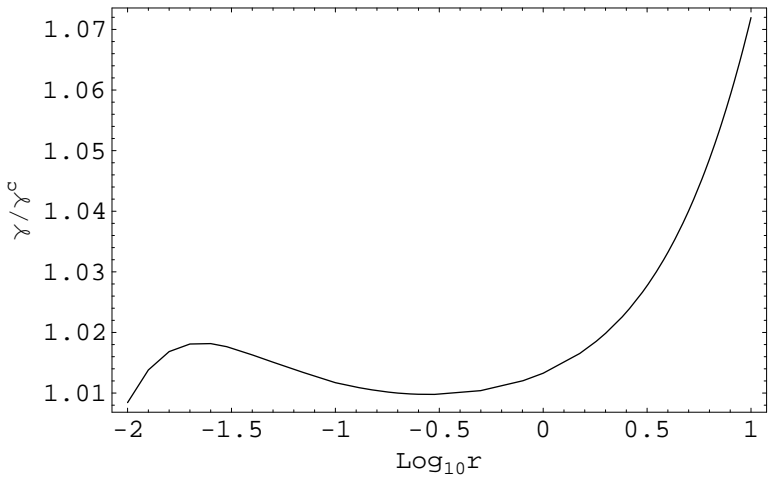}
 \caption{KH growth rate versus entropy ratio $r$ for standard neutron star parameters (Case A) and zero viscosity, normalised to the classical growth rate ($\gamma^{\textrm{c}}= 1.77\times10^{15}$ s$^{-1}$ at $U=10^6$ m s$^{-1}$). The
growth rate satisfies $\gamma\propto U^3 \sigma^{-1}$.}
 \label{fig7}
\end{figure}

\begin{figure}
 \includegraphics{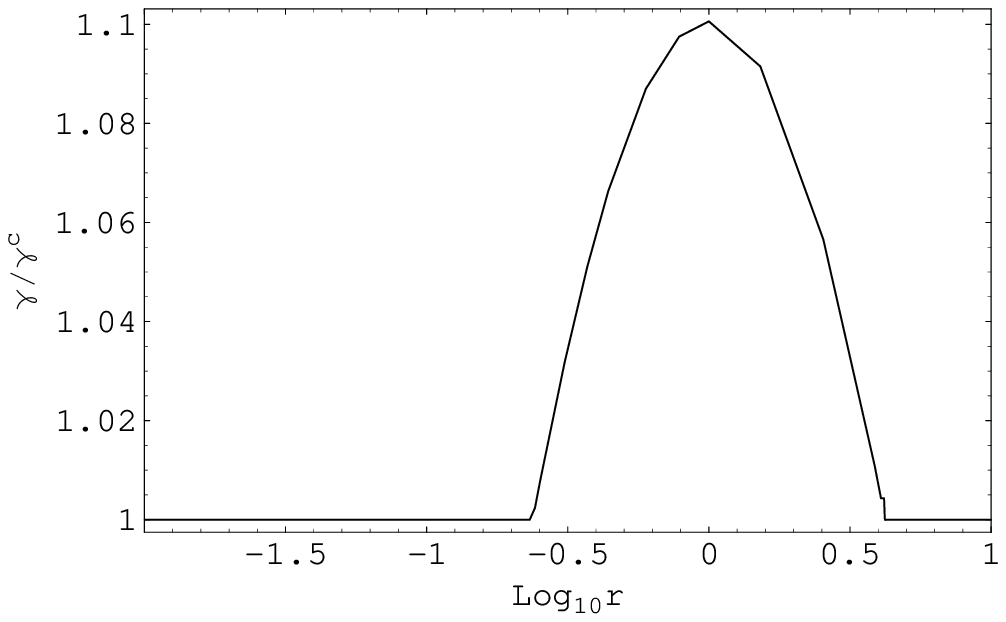}
 \caption{KH growth rate versus entropy ratio $r$ for standard neutron star parameters (Case A), viscosity $\nu=0.1$ m$^2$ s$^{-1}$, and zero surface tension, normalised to classical growth rate ($\gamma^{\textrm{c}}= 2.27 \times 10^{12}$ s$^{-1}$). The growth rate satisfies $\gamma \propto U^2 \nu^{-1}$.}
 \label{fig8}
\end{figure}


\appendix

\section{Integrating over Interfacial Discontinuities}

In general, a composite function of the form $g(x)f'(x)$, where
both $f(x)$ and $g(x)$ are discontinuous at an interface $x=0$, cannot be integrated across the discontinuity. Essentially,
this is because the integral $\int dx\, \delta(x) \theta(x)$, where
$\delta(x)$ and $\theta(x)$ are the Dirac delta and Heaviside step
functions respectively, is ill-defined. However, if one assumes
that the discontinuous functions change smoothly over the {\it{same}} small
length-scale $\epsilon$ within the infinitesimal interval $\epsilon \leq x
\leq \epsilon$, with $\epsilon \rightarrow 0$, the integral can be
regularized. For example, to integrate $g(x)f'(x)$, one writes
$f=\frac{b-a}{2\epsilon}x+\frac{b+a}{2}$ and
$g=\frac{d-c}{2\epsilon}x+\frac{d+c}{2}$, such that the functions
take well-defined values at the endpoints $f(-\epsilon)=a$, $g(-\epsilon)=c$, $f(\epsilon)=b$, and $g(\epsilon)=d$, yielding

\begin{equation}\lim_{\epsilon\rightarrow 0}\int_{-\epsilon}^{+\epsilon}dx\,gf'=\frac{(b-a)(d+c)}{2}.\end{equation}

Another combination one encounters in (20) and (21) involves four
discontinuous functions, of the form $l(x)h(x)g(x)f'(x)$. One
writes $f=\frac{b-a}{2\epsilon}x+\frac{b+a}{2}$, $g=\frac{d-c}{2\epsilon}x+\frac{d+c}{2}$, $h=\frac{\beta-\alpha}{2\epsilon}x+\frac{\beta+\alpha}{2}$,
and $l=\frac{\delta-\gamma}{2\epsilon}x+\frac{\delta+\gamma}{2}$,
yielding

\[
\lim_{\epsilon\rightarrow 0}\int_{-\epsilon}^{+\epsilon}dx\,lhgf'\]
\begin{equation}
=\frac{1}{24}(b-a)\left[(\delta+\gamma)(\beta-\alpha)(d-c)+(\delta-\gamma)(\beta+\alpha)(d-c)+(\delta-\gamma)(\beta-\alpha)(d+c)\right]+\frac{1}{8}(b-a)(\delta+\gamma)(\beta+\alpha)(d+c).\end{equation}

One can replace the linear trial functions above with other smooth trial functions, such as
trigonometric functions; the results do not
change. We emphasize again that there is no unique result for integrals of
products of discontinuous functions of the kind above. For example, $\int
dx\, g(x)f'(x)$ takes different values depending on the relative
length-scales over which $g$ and $f$ change, which are determined physically. The results
above pertain to the case where both $g$ and $f$ jump over an
interval O$(\epsilon)$.

\bsp \label{lastpage}


\begin{thebibliography}{}

\bibitem[\protect\citeauthoryear{Andersson, Comer \& Glampedakis}{2004}]{52} Andersson  N., Comer G.L., Glampedakis K., 2004,
preprint(astro-ph/0411748)
\bibitem[\protect\citeauthoryear{Andersson, Comer \& Prix}{2003}]{55} Andersson  N., Comer G.L., Prix R., 2003,
Phys. Rev. Lett., 90, 091101
\bibitem[\protect\citeauthoryear{Barenghi \& Jones}{1988}]{40b} Barenghi C., Jones M., 1988,
Journal of Fluid Mech., 197, 551
\bibitem[\protect\citeauthoryear{Bartkowiak et al.}{2004}]{51} Bartkowiak M., Fisher S.N., Gu\'{e}nault A.M., Haley R.P., Pickett G.R., Skyba P., 2004,
Phys. Rev. Lett., 93, 045301
\bibitem[\protect\citeauthoryear{Baym \& Pethick}{1975}]{7} Baym G., Pethick C.J., 1975,
Ann. Rev. Nucl. \& Particle Phys., 1975, 27
\bibitem[\protect\citeauthoryear{Blaauwgeers et al.}{2003}]{4} Blaauwgeers R., Eltsov V.B., Eska G., Finne A.P., Haley R.P., Krusius M., Ruohio J.J., Skrbek L., Volovik G.E., 2002,
Phys. Rev. Lett., 89, 155301
\bibitem[\protect\citeauthoryear{Brand \& Pleiner}{1981}]{9} Brand H., Pleiner H., 1981,
Phys. Rev. D, 24, 3048
\bibitem[\protect\citeauthoryear{Chandrasekhar}{1961}]{39} Chandrasekhar S., 1961,
Hydrodynamic and Hydromagnetic Stability. Dover Publications Inc.,
New York
\bibitem[\protect\citeauthoryear{Flowers, Ruderman \& Sutherland}{1976}]{16} Flowers E., Ruderman M., Sutherland P., 1976,
ApJ, 205, 541
\bibitem[\protect\citeauthoryear{Hall \& Vinen}{1956}]{41} Hall H.E., Vinen W.F., 1956,
Proc. Roy. Soc. Lond. A, 238, 215
\bibitem[\protect\citeauthoryear{Hoffberg et al.}{1970}]{25} Hoffberg M, Glassgold A.E., Richardson R.W., Ruderman M., 1970,
Phys. Rev. Lett., 24, 775
\bibitem[\protect\citeauthoryear{Jones}{1998}]{26} Jones P.B., 1998,
MNRAS, 296, 217
\bibitem[\protect\citeauthoryear{Khalatnikov}{2000}]{32} Khalatnikov I.M., 2000,
An Introduction to the Theory of Superfluidity. Perseus
Publishing, Cambridge
\bibitem[\protect\citeauthoryear{Khodel, Khodel \& Clark}{2001}]{27} Khodel V.V., Khodel V.V., Clark J.W., 2001,
Nucl. Phys. A, 679, 827
\bibitem[\protect\citeauthoryear{Korshunov}{2002}]{57} Korshunov S.E., 2002,
JETP Letters, 75, 423
\bibitem[\protect\citeauthoryear{Landau}{1941}]{34} Landau L.D., 1941,
J. Phys., V, 71
\bibitem[\protect\citeauthoryear{Landau \& Lifshitz}{1987}]{30} Landau L.D., Lifshitz E.M., 1987,
Fluid Mechanics, 2nd ed. Pergamon Press, Oxford
\bibitem[\protect\citeauthoryear{Lee}{1997}]{38} Lee D.M., 1997,
Rev. Mod. Phys., 69, 645
\bibitem[\protect\citeauthoryear{Leggett}{1975}]{35} Leggett A.J., 1975,
Rev. Mod. Phys., 47, 331
\bibitem[\protect\citeauthoryear{Link, Epstein \& Baym}{1993}]{56} Link B., Epstein R.I, Baym G., 1993,
ApJ, 403, 285
\bibitem[\protect\citeauthoryear{Mendell}{1991}]{60} Mendell G., 1991,
ApJ, 380, 515
\bibitem[\protect\citeauthoryear{Nomoto \& Tsuruta}{1987}]{21} Nomoto K., Tsuruta S., 1987,
ApJ, 312, 711
\bibitem[\protect\citeauthoryear{Osheroff \& Cross}{1977}]{50} Osheroff D.D., Cross M.C., 1977,
Phys. Rev. Lett., 38, 905
\bibitem[\protect\citeauthoryear{Privorotskii}{1975}]{53} Privorotskii I.A., 1975,
Phys. Rev. B, 12, 4825
\bibitem[\protect\citeauthoryear{Putterman}{1974}]{29} Putterman S.J., 1974,
Superfluid Hydrodynamics. North-Holland Publishing Company,
Amsterdam
\bibitem[\protect\citeauthoryear{Svidzinsky}{2003}]{13} Svidzinsky A.A., 2003,
ApJ, 590, 386
\bibitem[\protect\citeauthoryear{Shemar \& Lyne}{1996}]{78} Shemar S.L., Lyne A.G., 1996,
MNRAS, 282, 677
\bibitem[\protect\citeauthoryear{Tamagaki}{1970}]{6} Tamagaki R., 1970,
Prog. Theo. Phys., 44, 905
\bibitem[\protect\citeauthoryear{Tang et al.}{1991}]{54} Tang Y.H., Hahn I., Bozler H.M., Gould C.M., 1991,
Phys. Rev. Lett., 67, 1775
\bibitem[\protect\citeauthoryear{Tanigawa, Matsuzaki \& Chiba}{2004}]{80} Tanigawa T., Matsuzaki M., Chiba S., 2004,
Phys. Rev. C, 70, 065801
\bibitem[\protect\citeauthoryear{Tilley \& Tilley}{1994}]{31} Tilley D.R., Tilley J.,
1994, Superfluidity and Superconductivity, 3rd ed. IOP Publishing,
Bristol
\bibitem[\protect\citeauthoryear{Tsuruta}{1979}]{8} Tsuruta S., 1979,
Phys. Reports, 56, 237
\bibitem[\protect\citeauthoryear{Tsuruta}{1998}]{18} Tsuruta S., 1998,
Phys. Reports, 292, 1
\bibitem[\protect\citeauthoryear{Tsuruta et al.}{2001}]{17} Tsuruta S., Canuto V., Lodenquai J., Ruderman M., 1972,
ApJ, 176, 739
\bibitem[\protect\citeauthoryear{Vollhardt \& W\"{o}lfle}{2002}]{33} Vollhardt D., W\"{o}lfle P.,
2002, The Superfluid Phases of Helium 3. Taylor \& Francis, London
\bibitem[\protect\citeauthoryear{Volovik}{2002}]{40} Volovik G.E., 2002,
JETP Letters, 75, 418
\bibitem[\protect\citeauthoryear{Wheatley}{1975}]{36} Wheatley J.C., 1975,
Rev. Mod. Phys., 47, 415
\bibitem[\protect\citeauthoryear{W\"{o}lfle}{1979}]{37} W\"{o}lfle P., 1979,
Rep. Prog. Phys., 42, 271
Ann. Rev. Astron. Astrophys., 42, 169
\bibitem[\protect\citeauthoryear{Yakovlev, Levenfish \& Shibanov}{1999}]{10} Yakovlev D.G., Levenfish K.P.,
Shibanov Yu.A., 1999, Phys.Usp., 169, 825
\end{thebibliography}
\end{document}